\documentclass[prl, aps, 10pt, showpacs, superscriptaddress, twocolumn, floatfix]{revtex4-1}
\usepackage{graphicx}
\usepackage[usenames,dvipsnames]{color}
\usepackage{siunitx}
\usepackage{enumerate}
\usepackage{amsmath}
\usepackage{amssymb}
\usepackage{hyperref}
\usepackage{braket}

\usepackage{cleveref}

\definecolor{darkgreen}{rgb}{0.05,0.59,0.29}
\definecolor{orange}{rgb}{0.93, 0.53, 0.18}

\crefname{equation}{equation}{}
\Crefname{equation}{Equation}{}
\crefrangelabelformat{equation}{(#3#1#4--#5#2#6)}

\crefmultiformat{equation}{equations (#2#1#3}{, #2#1#3)}{#2#1#3}{#2#1#3}
\Crefmultiformat{equation}{Equations (#2#1#3}{, #2#1#3)}{#2#1#3}{#2#1#3}

\newcommand{\ff}{$|\mathcal{F}_{\mathrm{M0}^+}(Q^2)|^2$}

\newcommand{\he}{\mbox{$^4$He}}
\newcommand{\al}{\mbox{$^{27}$Al}}

\begin{document}
\title{Measurement of the $\alpha$-particle monopole transition form factor challenges theory: \\ a low-energy puzzle for nuclear forces?}



\def\kph{\affiliation{Institut f\"ur Kernphysik, Johannes
  Gutenberg-Universit\"at Mainz, D-55099 Mainz, Germany}}
\def\him{\affiliation{Helmholtz-Institut Mainz, Johannes Gutenberg-Universit\"at Mainz, D-55099 Mainz, Germany}}
\def\stefan{\affiliation{Jo\v zef Stefan Institute, 
    SI-1000 Ljubljana, Slovenia}}
\def\unil{\affiliation{Faculty of Mathematics and Physics, 
    University of Ljubljana, SI-1000 Ljubljana, Slovenia}}
\def\knt{\affiliation{University of Virginia, 
    Charlottesville, Virginia 22903, USA}}
\def\zagreb{\affiliation{Department of Physics, Faculty of Science, 
    University of Zagreb, 10000 Zagreb, Croatia}}
\def\cea{\affiliation{CEA IRFU/SPhN Saclay, F-91191 Gif-sur-Yvette
    Cedex, France}}
\def\clermont{\affiliation{Universit\'e Clermont Auvergne,
    CNRS/IN2P3, LPC, F-63000 Clermont-Ferrand, France}}
\def\racah{\affiliation{Racah Institute of Physics,
    Hebrew University, 91904 Jerusalem, Israel}}
\def\trento{\affiliation{Dipartimento di Fisica,
    Universit\`{a} di Trento, Via Sommarive 14, I-38123 Trento, Italy}}
\def\trentoTIFPA{\affiliation{Instituto Nazionale di Fisica Nucleare,
    TIFPA, Via Sommarive 14, I-38123 Trento, Italy}}

\author{S.~Kegel}\kph
\author{P.~Achenbach}\kph
\author{S.~Bacca}\kph\him
\author{N.~Barnea}\racah
\author{J.~Beri\v{c}i\v{c}}\stefan
\author{D.~Bosnar}\zagreb   
\author{L.~Correa}\clermont\kph
\author{M.\,O.~Distler}\kph
\author{A.~Esser}\kph
\author{H.~Fonvieille}\clermont
\author{I.~Fri\v{s}\v{c}i\'c}
\zagreb
\author{M.~Heilig}\kph
\author{P.~Herrmann}\kph
\author{M.~Hoek}\kph
\author{P.~Klag}\kph
\author{T.~Kolar}\unil\stefan
\author{W.~Leidemann}\trento\trentoTIFPA
\author{H.~Merkel}\kph
\author{M.~Mihovilovi\v{c}}\kph\stefan
\author{J.~M\"uller}\kph
\author{U.~M\"uller}\kph
\author{G.~Orlandini}\trento\trentoTIFPA
\author{J.~Pochodzalla}\kph
\author{B.\,S.~Schlimme}\kph
\author{M.~Schoth}\kph
\author{F.~Schulz}\kph
\author{C.~Sfienti}\kph
\author{S.~\v{S}irca}\unil\stefan
\author{R.~Spreckels}\kph  
\author{Y.~St\"ottinger}\kph
\author{M.~Thiel}\kph
\author{A.~Tyukin}\kph
\author{T.~Walcher}\kph
\author{A.~Weber}\kph

\date{\today}

\begin{abstract}
 We perform a systematic study of the $\alpha$-particle excitation from its 
ground state $0_1^+$ to the  $0_2^+$ resonance. The so-called monopole 
transition form factor is investigated via an electron scattering experiment in a broad $Q^2$-range 
(from $0.5$ to $5.0$ fm$^{-2}$).
The precision
of the new data dramatically superseeds that of older sets of data, each covering 
only a portion of the $Q^2$-range. The new data allow the 
determination of two coefficients in a low-momentum expansion leading to a new puzzle. By confronting
experiment to state-of-the-art theoretical  
calculations we observe that modern
nuclear forces, including those derived within chiral effective field theory 
which are well tested on a variety of observables, fail to reproduce 
the excitation of the $\alpha$-particle. 
\end{abstract}


\maketitle

The rather complex nature of the strong interaction generates a broad range of diverse phenomena in the Universe, that can be experimentally observed and theoretically interpreted. Among the phenomena that are most easily experimentally accessible are those involving strongly interacting matter in the form of atomic nuclei, built from quarks and gluons confined into nucleons.
At energies characteristic for nuclear binding the strength and complexity of quantum chromodynamics (QCD) complicates immensely the understanding of nuclear phenomena in terms of quarks and gluons as fundamental degrees of freedom. 
Providing a link between QCD with its inherent symmetries and the strong force acting in nuclear systems is a key problem in modern nuclear physics. From the theoretical point of view, a major breakthrough was spurred by the introduction of  the concept of effective field theory, which applied to low-energy QCD, gave rise to the so called interactions from chiral effective field theory ($\chi$EFT)~\cite{vankolck1994,bedaque2002,epelbaum2009,MACHLEIDT}, where nucleon-nucleon (NN),  three-nucleon (3N) (and more-nucleon) forces arise in a natural  and consistent hierarchical scheme. Developments in $\chi$EFT, together with advancements  in few/many-body methods, enable controlled calculations of matter at nuclear densities, and recently  even offer the opportunity to extend into the high density regimes of nuclear matter found in neutron stars~\cite{Drischler2021}.


In order to understand exciting phenomena such as neutron star mergers~\cite{NSmerger}, knowledge of the nuclear equation of state  is required. The latter is described in terms of a few parameters, such as the symmetry energy, its slope and
the  incompressibility~\cite{STEINER}. These quantities,  ultimately stemming from QCD and recently derived from $\chi$EFT using ab-initio methods~\cite{hagen2015,Hu:2021trw}, can be connected to the physics of finite nuclei. For example, the  incompressibility $K$,
giving information about the stiffness of nuclear matter against variations in the density, has been traditionally extracted from studies of the isoscalar monopole resonance~\cite{Blaizot1980},  interpreted as a breathing mode.  

In this Letter, we study the isoscalar monopole resonance  of the $\alpha$-particle in its transition from the ground state $0^+_1$ to the first excited state $0^+_2$. The $\alpha$-particle
is particularly interesting for a number of reasons. First,  the
$\alpha$-particle is the first light nucleus which has a binding energy per nucleon that resembles that of heavier nuclei. Second, the
$\alpha$-particle can be studied as a
 four-nucleon problem with numerical methods that are accurate even at the sub-percent level, leaving any discrepancy with experiment to be blamed 
  on the only input, namely the assumptions on the used nuclear Hamiltonian, consisting in the choice of nucleons as effective degrees of freedom and of their relative effective potential.
Third,
the first excited state $0^+_2$ is located between the proton and neutron breakup threshold, making it a very special case.  
Fourth, a rather low  incompressibility was found in a study of this nucleus~\cite{Bacca_PRC91} using both $\chi$EFT and other realistic potentials.

\begin{figure}
  \begin{center}
    \includegraphics[angle=0, width=\columnwidth]{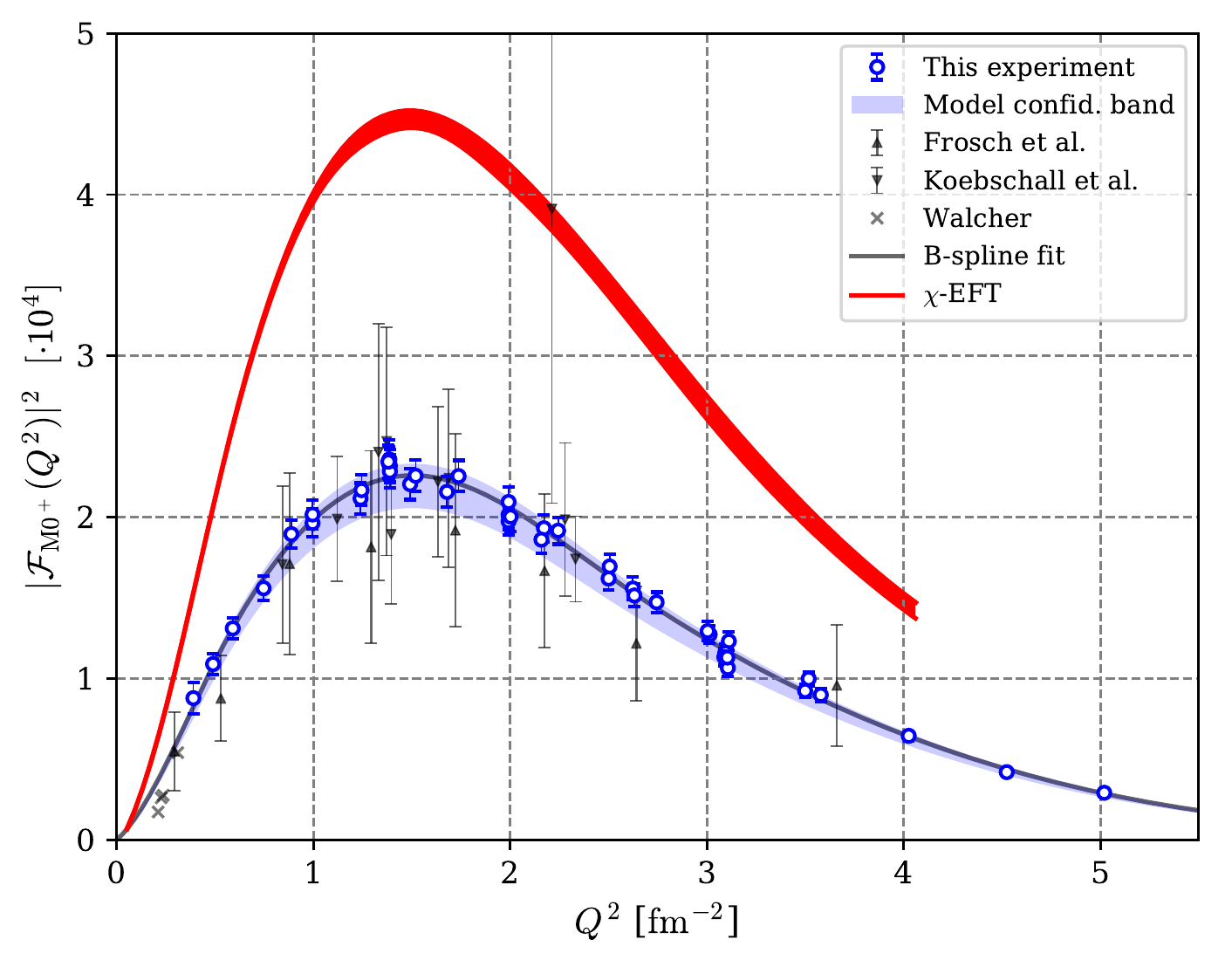}
    \caption{Monopole transition form factor as a function of $Q^2$, in comparison to previous data \cite{WalcherTh,KOEBSCHALL,FROSCH} and $\chi$EFT prediction~\cite{BaccaMono} (see text for details).
}
    \label{fig:Results}
  \end{center} 
\end{figure}

A further reason to study the $\alpha$-particle  
$0^+_1 \rightarrow 0^+_2$ transition is that a calculation based on $\chi$EFT NN+3N potentials~\cite{BaccaMono}
showed a strong disagreement with the existing experimental data ($\chi$EFT being up to a factor
 100$\%$ larger, see Fig.~\ref{fig:Results}), as opposed to an earlier calculation with a simple potential~\cite{Hiyama} which reproduced the data.
Two observations are in order: $(i)$ the existing data suffered from large uncertainties, and were taken at different facilities,  each covering only a portion of the interesting $Q^2$-range; 
$(ii)$ the available calculations were obtained with distinct few-body solvers, raising the doubt that differences may stem from the solver, not the Hamiltonian. This Letter aims at shedding light on both points.

 
 An extensive experimental campaign to measure the $^4$He monopole transition form factor with high precision was performed at the Mainz Microtron MAMI \cite{Herminghaus:1976mt} using the three spectrometer setup of the A1 collaboration \cite{Blomqvist98}. The continuous-wave electron beam with energies of \SI{450}, \SI{690}~ and \SI{795}{\mega\eV} was rastered on a cryogenic helium target. The target consisted of cryogenic helium gas encapsulated by an aluminium cell with \SI{250}{\micro\metre} thick walls. Permanent pressure and temperature measurements of the target gas allowed a precise determination of the helium density in the order of $\rho_{^4\mathrm{He}}=40~\mathrm{mg} / \mathrm{cm}^3$.  The two high resolution magnet spectrometers A and B were positioned at various forward scattering angles, covering the $Q^2$ range from \SI{0.5}{\per\square\femto\metre} to \SI{5.0}{\per\square\femto\metre}.\par
\noindent
The electrons were detected using four layers of vertical drift chambers (VDC) for particle track reconstruction, two layers of scintillators as trigger system, and a gas Cherenkov detector for electron-pion separation. The overall relative momentum resolution between $\SI{430}{\mega\eV}/\text{c}$ and $\SI{780}{\mega\eV}/\text{c}$ was determined to be $\delta \approx \num{2e-4}$. Due to the large relative momentum acceptances of spectrometer A (20 \%) and B (15 \%), both the monopole resonance and the \he{} elastic peak could be detected simultaneously.
\noindent
From the \he{} elastic peak and its corresponding form factor \cite{OTTERMANN}, the experimental luminosity can be determined precisely. In addition, the width of the elastic peak is used to estimate the experimental resolution needed for precise extraction of the width $\Gamma_0$ of the monopole resonance. \par
\noindent
In the analysis of the data to obtain the transition form factor, background determination and subtraction is an essential part. Background contributions originate from electrons scattering of the aluminium cell walls, the radiative tail of the elastic peak, and the continuum of \he{}. To determine the background contribution from the aluminium cell, almost exclusively consisting of \al{}, dedicated runs were performed for each setup with effectively reduced target gas density ($\rho_{\,^4\mathrm{He}}\approx 4\;\mathrm{mg}/\,\mathrm{cm}^3$). A measurement with a complete empty cell was not performed to avoid thermal stressing of the cell material. With these data, simulations for a complete model of the target cell background were designed and tested, including typical electron scattering processes on \al{}. Phenomenological models for elastic form factors \cite{FriedrichVoeglerHelm} were used to simulate the elastic scattering of \al{}, providing a good agreement between model and data taken with reduced target gas density.

\begin{figure}
  \begin{center}
    \includegraphics[angle=0, width=\columnwidth]{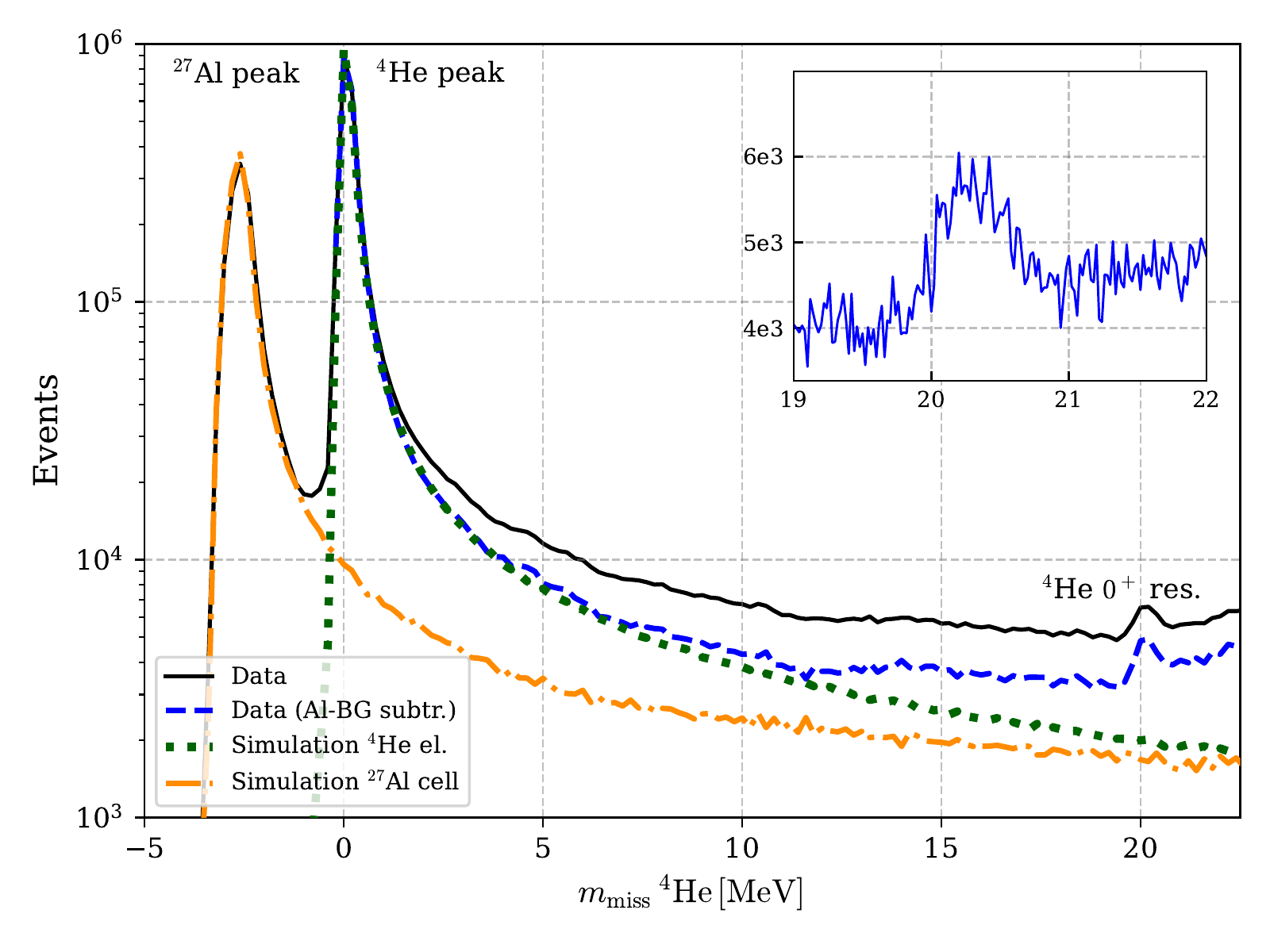}
    \caption{Missing mass spectrum of \he. The data is shown in blue (color online), in comparison to simulation of the \he{} elastic peak (green), and background contributions from the \al{} target. The contributions from the \al{} elastic peak is located at $m_\mathrm{miss} < 0$ due to recoil corrections applied to the scattered electrons. The insert shows the area of interest: the monopole resonance centered at \SI{20.21}{\mega\eV}.}
    \label{fig:MMissAll}
  \end{center}
\end{figure}

\begin{figure}
  \begin{center}
    \includegraphics[angle=0, width=\columnwidth]{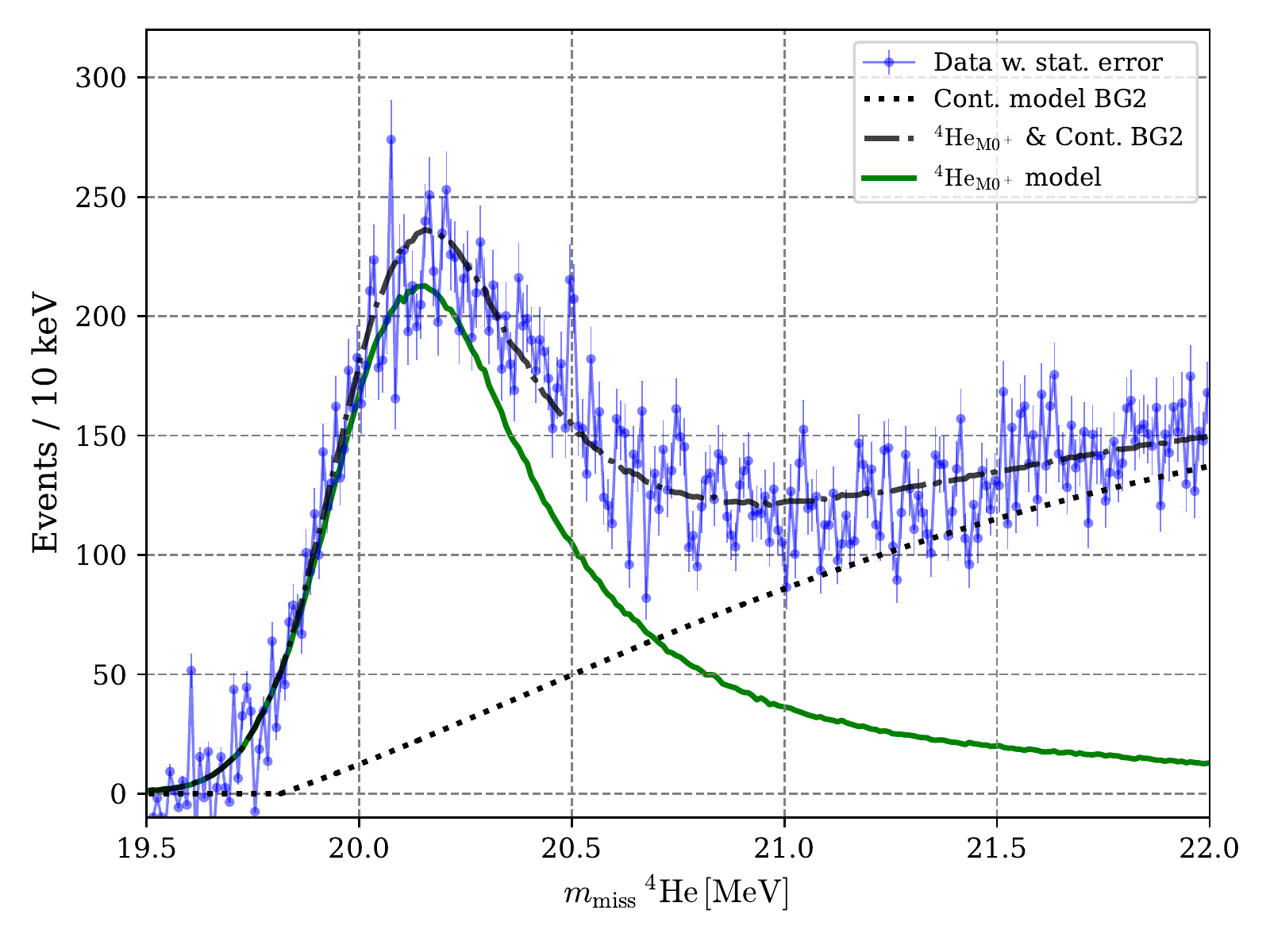}
    \caption{Typical $m_\mathrm{miss}$-mass spectrum of the monopole resonance of \he{}. Data (blue points-color online) in comparison to simulation of the monopole resonance (solid green online) and background model BG2 (black dotted line) based on the parametersiation of \cite{Jackson}. The dashed line shows the combination of background model and resonance simulation.}
    \label{fig:Monopol}
  \end{center}
\end{figure}

\noindent
Moreover, the excited states of \al{} listed in Ref.~\cite{NuDatAl27} could be observed in the missing mass ($m_\mathrm{miss}$) spectrum and were embedded in the simulation as well. 
An additional background contribution is given by two-body break-up processes from quasi-elastic electron scattering on \al{} for $m_\mathrm{miss} \gtrsim \SI{8}{\mega\eV}$ and thus propagate to the region of the monopole resonance. The break-up continuum of \al{} begins with the proton knock-out threshold $E_p=\SI{8.2}{\mega\eV}$ and is dominated by two-body break-up processes. Those processes are described and simulated as off-shell-electron scattering in the framework of DeForest \cite{DEFOREST1983232}. Furthermore, the form factor parameterisation from Ottermann et al.~\cite{OTTERMANN} has been used to simulate the additional source of background originating from the \he{} elastic peak. Radiative corrections, leading to a modification of the final electron energy to about 25 \%, were as well taken into account in all simulations \cite{VDHRadiative}. \autoref{fig:MMissAll} shows data and background simulations for the setup with \SI{450}{\mega\eV} beam energy at a scattering angle $\vartheta_\mathrm{scat.}=\SI{18.3}{\degree}$. After background subtraction, the \he{} continuum with the monopole resonance centered at \SI{20.21}{\mega\eV} between proton- and neutron-breakup threshold, is completely separated as illustrated in \autoref{fig:Monopol}.\par\noindent
To extract the transition form factor from the measured cross sections, an appropriate model parametrisation of the unknown resonance peak, including radiative corrections, is required. For such parameterisations, the resonance peak is traditionally considered as convolution of a Gaussian and a Lorentzian distribution, $G(E,E_0,\sigma_\mathrm{res.})$ and $L(E,E_0,\Gamma_0)$ respectively. While the width of the Gaussian distribution, $\sigma_\mathrm{}$, includes experimental resolution effects, the intrinsic width  of the resonance is implemented by the full-width-at-half-maximum (FWHM) $\Gamma_0$ of the Lorentzian distribution. To avoid the complicated convolution integral, the following  approximation is used
\begin{equation}
  \label{eq:Voigt}
    \sigma_\mathrm{1}(E,E_0,\sigma_\mathrm{res.},\Gamma_0) 
    \! \propto \! \eta  L(E,E_0,\Gamma_0) + 
    (1- \eta) G(E,E_0,\sigma)\;.
\end{equation}

\noindent
In this ansatz $\eta \in (0,1)$ is a parameter, constrained by $\sigma$ and $\Gamma_0$, which regulates the ratio of Gaussian to Lorentzian distribution \cite{VoigtLiu}. In order to quantify more precisely the systematic uncertainties of the described model, a second approach has been used as described in \cite{Jackson}, \cite{MarguilesPhelan}. Within this model, the resonance depends on two dimensionless parameters
\begin{equation}
  \label{eq:Jackson}
      \mu  =  \frac{E-E_\mathrm{thresh.}}{\Gamma_0 /2} {~\rm and~}
       \mu_0  =  \frac{E_0-E_\mathrm{thresh.}}{\Gamma_0 /2}
       \end{equation}
     with $E_0=20.21$ MeV being the central value of the resonance and $E_\mathrm{thresh.}=19.815$ MeV the continuum proton-threshold. The resonance parameterisation is taken as
 \begin{equation}
    \label{eq:Jacksonc}
    \sigma_\mathrm{2}(\mu, \mu_0)  \propto  \frac{(\mu / \mu_0)^{\frac{1}{2}}}{ (\mu-\mu_0)^{2} + (\mu / \mu_0)} .
\end{equation}

\noindent
As consequence of \cref{eq:Jackson,eq:Jacksonc}, no resonance
events appear below this threshold, as long as resolution and radiative effects are neglected. Resolution effects are implemented in this parameterisation by an additional uncertainty of the momentum and angular reconstruction of the spectrometers. This uncertainty is represented by a superposition of two Gaussian distributions with different widths.
\noindent
To match simulation and data, a complete determination of the previously unknown parameters $\sigma_\mathrm{res.}$ and $\Gamma_0$ is mandatory. The experimental resolution for both parametersations is determined by the width of the \he{} elastic peak, which is broadened by resolution effects as well as radiative losses. These two contributions are disentangled by Monte-Carlo techniques and verified by data. In order to describe the background contribution from the \he{} continuum two model descriptions have been applied in order to quantify model uncertainties. One model (BG1) describes the continuum under the resonance peak as a linear function, while the second model (BG2) is based on the assumption that the resonance is located on the left tail of a broad giant resonance at \SI{25.95}{\mega\eV} with $I^\mathrm{P}=1^-$ \cite{Tilley}. For the determination of $\Gamma_0$, the simulations of Eq.~\eqref{eq:Voigt} and Eq.~\eqref{eq:Jacksonc} as well as the two background models were compared to data with $\Gamma_0$ as free parameter to be optimised.
\begin{table}[h!]
  \caption{FWHM $\Gamma_0$ for the investigated resonance parameterisations $\sigma_1$ Eq.~\eqref{eq:Voigt} and $\sigma_2$ Eq.~\eqref{eq:Jacksonc} and the two background parameterisations BG1 and BG2.}  \label{tab:gamma}
  \begin{ruledtabular}
  \begin{tabular}{ccc}
    -   & BG1               &  BG2 \\ \hline
    $\sigma_1$    & 268 $\pm$ 43 keV  & 285 $\pm$ 33 keV  \\
    $\sigma_2$    & 262 $\pm$ 47 keV  & 288 $\pm$ 39 keV  \\ \hline\hline 
  \end{tabular}
\end{ruledtabular}
\end{table}
Our results for $\Gamma_0$, summarised in \autoref{tab:gamma}, agree within 
error bars with previous data from Walcher \cite{WalcherTh_1,WalcherTh} and K\"obschall \cite{KOEBSCHALL}, while they disagree with data from Frosch~\cite{FROSCH},
and can be compared to the only theoretical calculation of
$^4$He$(e,e')$, which resulted in
$\Gamma_0=180(70)$ keV, using a central NN force~\cite{Winfried}.


The transition form factor is obtained from the experimental cross section divided by the normalised Mott cross section,

\begin{equation}
  \label{eq:Formfactor}
     |\mathcal{F}_{\mathrm{M0}^+}(Q^2)|^2 = \left(\frac{d\sigma}{d\Omega}\right)_\mathrm{Exp.} \bigg/ \;  \left( 4\pi \frac{d\sigma}{d\Omega}\right)_\mathrm{Mott}
\end{equation}

\noindent
where $Q^2=q^2-E_0^2$.
It is beneficial to take advantage of the simultaneously measured elastic peak of \he{} to avoid fluctuations in the data caused by different luminosities, and determine the monopole form factor relative to the elastic peak.
Both quantities, the elastic peak and the monopole resonance, exhibit a slightly different $Q^2$ which was accounted for when evaluating the form-factor ratio. The value of $Q^2$ is determined by a binned distribution taking into account the applied data cuts. Those cuts were first restricted to $\pm \SI{240}{\kilo\eV}$ around both peaks to keep the influence of the continuum background to the form-factor ratio small. The relative transition form factor established this way is then in a last iteration step improved by extending the $m_\mathrm{miss}$-cut from \SI{19.5}{\mega\eV} to \SI{22}{\mega\eV} to include large contributions of the monopoles radiative tail. For this purpose the resonance peak is simulated with parameterisation $\sigma_1$ and $\sigma_2$ and the valid transition form-factor ratio and in combination with the backgrounds BG1 and BG2 respectively optimised to data in order to minimise the $\chi^2$. Within this minimisation procedure the simulation of the monopole resonance peak is allowed to float only by a factor, which is then used to adjust the transition form factor.

Our final experimental results for the monopole transition form factor are shown in \autoref{fig:Results}, in comparison to the $\chi$EFT calculation from Ref.~\cite{BaccaMono}.
A third order basis spline polynomial is used to fit the data. To account for the model uncertainties, the analysis was repeated with all remaining combinations of resonance parameterisations and background models. Analysing the transition form factor with model BG1 leads to a variation of $\delta_\mathrm{BG\;model} = \pm 3.2 \%$ around the results obtained by BG2, \autoref{tab:syserrors}. However, a constant shift of the transition form factor to higher or lower values by a different continuum model could not be verified. On the contrary, analysing the data with $\sigma_1$ from \eqref{eq:Voigt} leads to an average shift of the transition form factor of $\delta_\mathrm{Res.\;model} = - 5.8 \%$ and thus to smaller values. These model dependencies were added linearily to the (blue)  model confidence band in \autoref{fig:Results}, representing the model uncertainty of the data.
The contributions to the total systematic uncertainty on the extraction of the transition form factor are summarised in \autoref{tab:syserrors}. The uncertainty of the elastic form factor of \he{}, used to normalise the data, has been estimated to 0.5 \% as given by the authors in \cite{OTTERMANN}. Background subtraction of the elastic tails from $^4$He, $^{27}$Al and the quasi elastic scattering off $^{27}$Al contribute to the systematic uncertainty with up to 1 \%. The FWHM of the monopole resonance $\Gamma_0$ influences the transition form factor \ff~by 4 \% and contributes the major uncertainty. This uncertainty has been estimated by varying $\Gamma_0$ within a realistic  error range and observing the effect onto the transition form factor. All systematic errors were added quadratically to the statistical errors.
Our results agree with  previous data~\cite{FROSCH,KOEBSCHALL} albeit have a much higher precision and thereby reinforce the tension with ab-initio calculations~\cite{BaccaMono}, where for example $\chi$EFT result is 
100$\%$ too high at $Q^2=1.5$ fm$^{-2}$ with respect to the new data.

\begin{table}[h!]
  \caption{Contributions to the systematic uncertainties of the transition form factor and the model dependencies.}\label{tab:syserrors}
  \begin{ruledtabular} 
  \begin{tabular}{ll}
    Source & $\Delta |\mathcal{F}_{\mathrm{M0}^+}(Q^2)|^2$\\ \hline 
    Background & $\pm 1\,\%$\\
    \he{} g.s. form factor & $\pm 0.5\,\%$\\ 
    $\Delta \Gamma_0$ & $\pm 4\,\%$\\ \hline\hline
    \multicolumn{2}{c}{Model uncertainties} \\ \hline\hline
    BG1-BG2               & $\pm 3.2\,\%$\\
    $\sigma_1$-$\sigma_2$ & $- 5.8\,\%$\\
  \end{tabular}
\end{ruledtabular}
\end{table}

\begin{figure}
  \begin{center}
    \includegraphics[angle=0, width=\columnwidth]{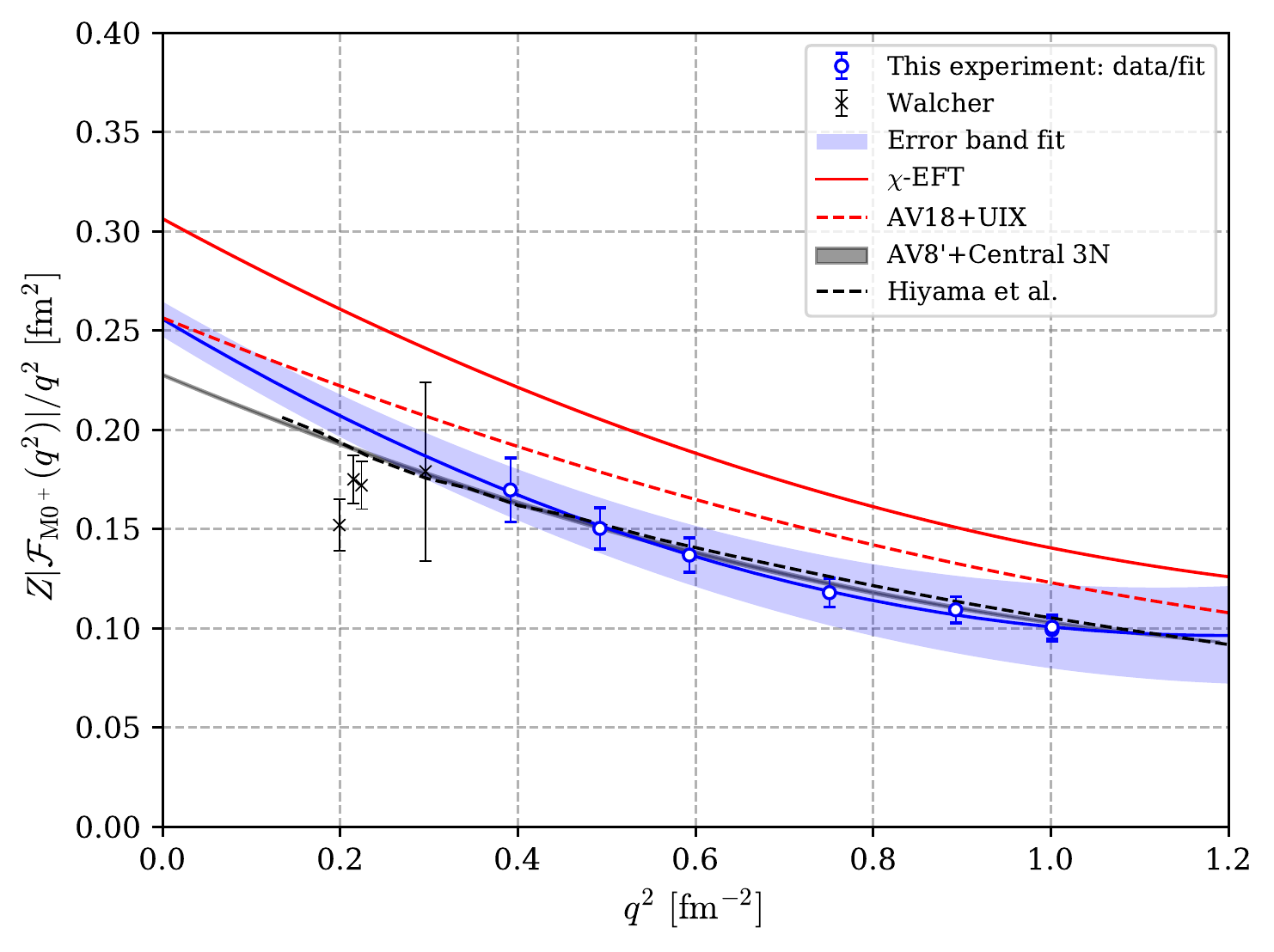}
    \caption{Low-$q$ data for the monopole form factor: theory vs. experiment (see text for details). 
    }
    \label{fig:radiusrange1}
  \end{center}
\end{figure}

One could argue that at high momentum transfer values $\chi$EFT is not expected to work well. Hence, we now focus on discussing the low-momentum region.
A $q\rightarrow0$ expansion, where $q^2$ is the three-vector momentum, yields~\cite{Theissen, chernykh} 
\begin{equation}
  \label{eq:expansion}
  \frac{Z |\mathcal{F}_{\mathrm{M0}^+}(q^2)|}{q^2} = \frac{\braket{r^2}_{tr} }{6} \left[ 1-\frac{q^2}{20}{\mathcal{R}^2}_\mathrm{tr} 
    + \mathcal{O}(q^4) \right] 
\end{equation}
 and allows to extract the monopole transition matrix element $ \braket{r^2}_{tr} $
and the transition radius $\mathcal{R}^2_{tr}=\braket{r^4}_{tr}/\braket{r^2}_{tr} $, which provide information about the spatial structure of the resonant state $0^+_2$. 
We use this formula to extract these quantities both from experimental data and theoretical calculations using a three-parameter fit.
For  the theory,  given that $\mathcal{F}_{\mathrm{M0}^+}$ was calculated on a grid of $q$ every 0.25 fm$^{-1}$~\cite{BaccaMono},
we use four available low-momentum points at $q^2 \leq 1$ fm$^{-2}$, assigning to each point a  1$\%$ numerical uncertainty. The fit values for $ \braket{r^2}_{tr}$ 
and $\mathcal{R}_{tr}$  are compatible with what we obtain from a direct
calculation using the transition density in coordinate space~\cite{Bacca_PRC91}. 
For the experiment, we fit the six  data points below $q^2=1$ fm$^{-2}$ neglecting the recoil and assuming a sharp resonance. 
The obtained values for $ \braket{r^2}_{tr} $ and $\mathcal{R}_{tr}$  are reported in \autoref{tab:radius} with the uncertainties given by the fit and the corresponsing curves based on the mean value of \autoref{tab:radius} are shown in Fig.~\ref{fig:radiusrange1}. 

We notice that in the range of $0.2 \leq q^2 \leq 1$ fm$^{-2}$ the simplified potential used by Hiyama~\cite{Hiyama} leads to agreement with
the experimental data, while the realistic calculations do not. Because the calculation by Hiyama was performed with a distinct few-body method, in this paper we recalculate it with the same method as in Ref.~\cite{BaccaMono}, so that a comparison with other interactions can be done on equal footing. In Fig.~\ref{fig:radiusrange1}
we show that we (grey solid line) reproduce the result of Ref.~\cite{Hiyama}  (black dashed line). We assign a 1$\%$ uncertainty to our calculation  by taking the difference from the largest and second largest model-space results. While describing the data, the AV8'+central 3N potential
 is however not compatible with the experimental fit
value of $\braket{r^2}_{tr}$, while the realistic AV18+UIX is.
Overall, we see that theory predicts a smaller value of 
${\mathcal R}_{tr}$ than the experimental fit, and the $\chi$EFT prediction deviates the most from  experiment, even at low-momenta.

The combination of the new experimental data and calculations prove that there is a puzzle which  is not due to the applied few-body method,
but rather to the modeling of the nuclear Hamiltonian.
 Interestingly, another recent investigation~\cite{Viviani}  shows that the $0^+_2$ state in $^4$He is very sensitive to the particular parameterization of the chiral 3N force.
 This puzzle may resemble other ---still unresolved---low-energy puzzles, such as 
the so called $A_y$~\cite{Ay_3b_2002,Golak,Deltuva,Pisa2019} and  SST~\cite{SST} puzzles, where the most sophisticated nuclear forces
are not able to explain experimental observations in the few-body sector.

Further theoretical work is needed to resolve the $\alpha$-particle monopole puzzle. An order-by-order calculation in $\chi$EFT would certainly be important to better assess the theoretical uncertainties which go beyond the numerical error. Furthermore, a variation of the low-energy constants in the three-body force could  shed light on the effect that a modification of the short-range dynamics has on the monopole transition. Work in this directions has been started~\cite{Simone}.
A systematic experimental verification of the future theoretical developments
will be opened up by the low-energy electron beam of the new Mainz Energy-recovering Superconducting Accelerator (MESA) under construction at Mainz~\cite{Mesa}, which will operate in the ideal energy regime to test $\chi$EFT. In the low-$q$ regime the investigation of the $\alpha$-particle monopole form factor will shed light on the apparent tension between the data of Ref.~\cite{WalcherTh} and the extrapolation based on this experiment performed at higher momenta.

\begin{table}
  \caption{Values of $ \braket{r^2}_{tr} $ and $\mathcal{R}_\mathrm{tr}$: experiment vs theory.} \label{tab:radius}
  \begin{ruledtabular} 
  \begin{tabular}{lll}
   &       $\braket{r^2}_{tr}$  [fm$^2$]           &  $\mathcal{R}_\mathrm{tr}$ [fm] \\
    \hline \\
  Experiment &  1.53 $\pm$ 0.05   &  4.56 $\pm$ 0.15 \\
  Theory (AV8'+ central 3N) & 1.36 $\pm$ 0.01 & 4.01 $\pm$ 0.05\\
  Theory (AV18+UIX) &  1.54 $\pm$ 0.01& 3.77 $\pm$ 0.08 \\
  Theory ($\chi$EFT) &  1.83 $\pm$ 0.01& 3.97 $\pm$ 0.05\\
  \hline
  \end{tabular}
\end{ruledtabular}
\end{table}

\noindent

\begin{acknowledgments}
\noindent
We acknowledge  support from the technical staff at the Mainz Microtron and thank the accelerator group for the excellent beam quality.
 S. B. would like to thank Alejandro Kievsky and Michele Viviani for useful discussion.  
This work was supported by the Deutsche Forschungsgemeinschaft (DFG) with the Collaborative Research Center 1044 and
through the Cluster of Excellence ``Precision Physics, Fundamental
Interactions, and Structure of Matter" (PRISMA$^+$ EXC 2118/1) funded by the
DFG within the German Excellence Strategy (Project ID 39083149). We also acknowledge the Israel Science Foundation grant number 1308/16. 
Theoretical calculations were run on the supercomputer MogonII at Johannes Gutenberg-Universit\"{a}t Mainz.
\end{acknowledgments}

\bibliographystyle{apsrev4-1} 
\bibliography{mono}
\end{document}